\documentclass[a4paper,12pt]{article}
\pdfoutput=1
\usepackage{hyperref}
\usepackage{epsfig}
\usepackage{amssymb}
\usepackage{cancel}
\usepackage{setspace}
\usepackage[usenames,dvipsnames,svgnames,table]{xcolor}
\usepackage{amsmath}
\usepackage{graphicx}
\usepackage{subfigure}
\usepackage{url}
\usepackage{slashed}
\usepackage{booktabs}
\newcommand{\ie}{{\it i.e.}}

\newcommand{\be}{\begin{equation}}
\newcommand{\ee}{\end{equation}}
\newcommand{\br}{\begin{eqnarray}}
\newcommand{\bea}{\begin{eqnarray}}
\newcommand{\eea}{\end{eqnarray}}
\newcommand{\er}{\end{eqnarray}}
\newcommand{\ba}{\begin{array}}
\newcommand{\ea}{\end{array}}
\newcommand{\bi}{\begin{itemize}}
\newcommand{\ei}{\end{itemize}}
\newcommand{\bn}{\begin{enumerate}}
\newcommand{\en}{\end{enumerate}}
\newcommand{\bc}{\begin{center}}
\newcommand{\ec}{\end{center}}

\newcommand{\beq}{\begin{equation}}
\newcommand{\eeq}{\end{equation}}

\newcommand{\gsim}{\lower1.0ex\hbox{$\;\stackrel{\textstyle>}{\sim}\;$}}
\newcommand{\lsim}{\lower1.0ex\hbox{$\;\stackrel{\textstyle<}{\sim}\;$}}

\newcommand{\bs}{\begin{small}}
\newcommand{\es}{\end{small}}

\begin{document}
\thispagestyle{empty}
\begin{flushright}
HIP-2016-37/TH \\
\end{flushright}
\vspace*{10mm}
\begin{center}
  {\LARGE {\bf Dark-photon searches \\ 
      \vspace{0.3cm}
      via  $ZH$  production at $e^+e^-$ colliders
}}\\
\vspace*{1.5cm}
{\bf Sanjoy Biswas$^{a}$, Emidio Gabrielli$^{b,c,d}$,
 Matti Heikinheimo$^{e}$,  Barbara Mele$^{f}$}\\

\vspace{0.5cm}
{\it(a)   Korea Institute for Advanced Study, 
85 Hoegi-ro, Seoul 02455, Republic of Korea
\\
(b)  Dipartimento di Fisica, Theoretical section, Universit\`a di 
Trieste, \\ Strada Costiera 11, I-34151 Trieste, Italy
\\ 
(c) INFN, Sezione di Trieste, Via Valerio 2, I-34127 Trieste, Italy
\\
(d) NICPB, R\"avala 10, Tallinn 10143, Estonia 
\\
(e) Helsinki Institute of Physics, University of Helsinki,
P.O. Box 64, Helsinki FI-00014, Finland
\\
(f) INFN, Sezione di Roma,  P. le A. Moro 2, I-00185 Rome, Italy
\\[1mm] }

\vspace*{2cm}{\bf ABSTRACT}
\end{center}
\vspace{0.3cm}

\noindent
We study the $ZH$  associated production followed by  the Higgs $H\to \gamma \bar{\gamma}$ decay into a photon plus an {\it invisible}  and {\it massless}  dark photon,  
at future high-energy $e^+e^-$ facilities. 
Large  \mbox{$H\to \gamma \bar{\gamma}$} decay rates (with branching ratios up to a few percent)   are allowed, thanks to possible non-decoupling properties of the Higgs boson under specific conditions, and unsuppressed dark-photon couplings in the dark sector. Such large decay rates can  be obtained in the framework of recent  flavor models that aim to naturally explain the observed spread in the fermion mass spectrum. 
 We analyze  the experimental prospects for observing the  $e^+e^-\rightarrow ZH$ process followed by the semi-invisible Higgs decay   into a photon plus a massless invisible system. Search strategies for both the leptonic  and the hadronic  final states  (arising from $Z\rightarrow \mu^+\mu^-$ and $Z\rightarrow q\bar{q}$, respectively) are outlined. We find that a $5\sigma$ sensitivity to a branching fraction $BR_{\gamma\bar{\gamma}}\sim 3\times 10^{-4}$ can be achieved by combining  the two channels with an integrated luminosity of 10 ab$^{-1}$ at a  c.m. energy
 of  240~GeV. This is considerably better than the corresponding sensitivity in alternative channels previously studied  at
 lepton  colliders. The analysis is model independent, and its results can be straightforwardly applied to the search of any Higgs two-body decay into a photon plus an undetected
light particle.
\newpage

\section{Introduction}
The Higgs-boson discovery at the LHC in 2012 \cite{Aad:2012tfa} marked a milestone in our understanding of the electroweak symmetry breaking via the Higgs-Englert-Brout mechanism \cite{Englert:1964et}. Present data are well consistent with the Standard Model (SM) expectations for the Higgs boson properties~\cite{Khachatryan:2016vau}, although there is still room, especially in the Higgs sector, for potential New Physics (NP) effects,  which could be detected in 
the forthcoming collider physics program. NP could for instance affect
 the chiral symmetry breaking, which is parametrised in the SM by the Higgs Yukawa couplings to fermions, and is responsible for the fermion mass spectrum, flavor mixing and CP violating phenomena, whose pattern is presently in excellent agreement with experiments.
Despite that, the  origin of Yukawa couplings is actually a mystery. Their eigenvalues span over six orders of magnitude for charged fermions and even more if neutrinos have Dirac masses. Such unexplained wide range of masses is often referred to as the flavor hierarchy problem. Indeed, 
it is not yet clear  whether the Yukawa couplings are fundamental constants (like gauge couplings), arising for instance from a ultraviolet (UV) completion of the SM or 
are just low-energy effective couplings. Although the latter possibility is presently the most promising to  explain the origin of the fermion mass hierarchy, it could require the existence of a non-trivial NP structure able to give rise to the effective Yukawa couplings. For instance, hidden or dark sectors beyond the SM could do the job, by promoting the Higgs boson to the role of a portal to the dark sector.

On the other hand, general consents are growing around the idea that a dark sector, weakly coupled to the SM, could be responsible for the observed dark matter (DM) in the Universe \cite{Ade:2013zuv,Ade:2015xua}. 
The dark-sector 
internal structure and interactions could include  light or massless $U(1)$ gauge bosons (the dark photons) which mediate long-range forces between dark particles \cite{Ackerman:mha,ArkaniHamed:2008qn,Fan:2013tia,Heikinheimo:2015kra}. In cosmology, dark photons may help to solve the problems related to the small-scale structure formation \cite{ArkaniHamed:2008qn}, and, if massless,  they can predict dark discs of galaxies~\cite{Fan:2013tia}. 
On the theoretical side, scenarios with dark (or hidden) photons  have been extensively investigated in the literature (especially in the framework of UV completions of the SM theory), both for massive and massless dark photons \cite{Holdom:1985ag}, \cite{Abel:2008ai}. This  has also motivated dedicated experiments \cite{Abazov:2009hn},  mainly focused on massive dark-photon searches though~\cite{Essig:2013lka}.
Recently, there has been a renewed interest for viable cosmological scenarios with 
DM  that  is  charged  under  a $U(1)$  gauge  group  in the  dark  sector,  decoupled  from  SM  forces, and mediated by massless dark photons~\cite{Agrawal:2016quu}. 
Constraints on DM charged under $U(1)$ interactions have been revisited, allowing for viable unexplored  cosmological models with large couplings in the dark sector.

A NP theoretical flavor framework, aiming to  solve not only the flavor hierarchy problem but also  the origin of DM, has been proposed in \cite{Gabrielli:2013jka}. The model can predict  an exponential spread in effective Yukawa couplings, and is based on an unbroken $U(1)$ gauge symmetry in a dark sector, providing a theoretical explanation for the existence of long-range dark interactions, as  suggested by cosmological observations~\cite{Moore:1994yx,Klypin:1999uc,Spergel:1999mh}.
The dark sector of the model contains  a set of massive dark fermions (heavier  SM-fermion replicas), which are SM singlet but are charged under the dark $U(1)$ gauge group. Furthermore,  heavy messenger scalar fields, charged under both  the dark $U(1)$ and the SM gauge group, 
are needed  to transfer at one loop the flavor and chiral symmetry breaking from the dark sector to the SM fermions. Incidentally, although the theory is not supersymmetric, 
the messenger fields have the same SM quantum numbers as squarks and sleptons in minimal supersymmetric models~\cite{Haber:1984rc}.

The main paradigm of the Gabrielli-Raidal flavor model (GRFM) in  \cite{Gabrielli:2013jka}   is that the Yukawa couplings are, rather than fundamental constants, effective low-energy couplings  generated radiatively by the interactions in the hidden sector of the theory. 
In particular, Yukawa couplings are assumed to vanish at  tree level by 
some symmetry (for a gauge-symmetry realisation, see~\cite{Gabrielli:2016vbb}), and are induced at one loop by  dark-sector fields~\cite{Gabrielli:2013jka}. Due to chirality, Yukawa couplings follow the dark-fermion mass hierarchy, which in the GRFM is exponential. 
Indeed, the dark-fermion exponential spectrum is generated by a non-perturbative dynamics in the dark sector involving $U(1)$ gauge interactions. Then, since the $U(1)$ gauge symmetry is exact, the dark fermions have to be stable, and therefore are potential DM candidates. Then, the GRFM can provide a basis for a viable charged DM scenario, as, for instance, the one suggested in \cite{Agrawal:2016quu}.

We stress that in the GRFM 
the observed quark and lepton spectrum can be  reproduced up to a few percent by  the exponential-spread relation for the dark-fermion masses~\cite{Gabrielli:2013jka,Gabrielli:2016vbb,Biswas:2015sha},
provided  dark-fermion $U(1)$ charges of the same order are assumed. Moreover,
the corresponding $U(1)$ fine structure constant can be predicted 
from the  lepton mass-spectrum sum rules  to be quite strong, although still within the perturbative regime~\cite{Gabrielli:2013jka,Gabrielli:2016vbb,Biswas:2015sha}.
Notice that one is indeed allowed to have a strongly coupled dark photon in the dark sector  only for massless dark photons, which can be fully decoupled at tree level from the SM quark and lepton sector~\cite{Holdom:1985ag}. In fact, most of present astrophysical and accelerator constraints apply to {\it massive} dark-photon couplings \cite{Abazov:2009hn}, for which {\it unavoidable} tree-level dark-photon couplings to SM matter fields arise~\cite{Holdom:1985ag}. 

Although it can be fully decoupled at tree level from
 SM particles, a massless dark photon can still have effective low-energy interactions with SM fields arising from higher dimensional operators, with the latter suppressed by a characteristic scale related to the mass of the messenger fields running in the loops. 
 For example, a massless dark photon ($\bar{\gamma}$) can appear in the flavor changing neutral current (FCNC) $f\to f^{\prime} \bar{\gamma}$ decays of the SM fermions~\cite{Gabrielli:2016cut}, that are mediated by FCNC 
 magnetic-dipole-type operators suppressed by the NP scale running in the loop. 
 
 On the contrary, dark-photon couplings to the Higgs boson can show non-decoupling properties (a typical example is  when the messenger fields have the same quantum numbers as squarks and sleptons~\cite{Gabrielli:2014oya}). 
An effective gauge-invariant low-energy $H\gamma\bar\gamma$ interaction can indeed arise at one loop. This interaction is induced by a gauge-invariant dimension-5 operator, suppressed by an effective scale $\Lambda_{eff}$, according to
\bea
    {\cal L}&=& \frac{1}{\Lambda_{eff}} H F_{\mu\nu}\bar{F}^{\mu\nu}\, ,
\label{LH}
\eea
where $F_{\mu\nu}$ and $\bar{F}_{\mu\nu}$ are the field strengths of the photon and
dark photon, respectively.  The effective high-energy scale $\Lambda_{ eff}$, as defined in Eq.(8) of \cite{Gabrielli:2014oya}, is
  \bea
  \Lambda_{eff}&=&\frac{6\pi v}{R\sqrt{\alpha\bar{\alpha}}}
  \frac{1-\xi^2}{\xi^2}\, ,
\label{eq1}
  \eea
where $v$ is the SM Higgs vev, $\xi\equiv \Delta/\bar{m}^2$ is the mixing parameter,  $\Delta=v \mu$ is the off-diagonal term appearing in the left-right messenger square-mass matrix, $\bar{m}$ is the average messenger mass, and  $\alpha$
and $\bar{\alpha}$ are the electromagnetic and $U(1)$ dark fine-structure constants, respectively. $R$ is given by a product of quantum charges (see for instance Eq.(4) in  \cite{Gabrielli:2014oya} for notations). The scale $\mu$ is connected to the vev of a heavy singlet  scalar field needed to generate effective Yukawa couplings at 1-loop \cite{Gabrielli:2013jka}. Importantly, the $\xi$ parameter  can be viewed as a relative square-mass  difference of the messenger mass eigenstates running in the loop [$m_{\pm}^2=\bar{m}^2(1\pm\xi)$], and
  should be positive and limited by 1, in order to avoid tachions in the spectrum. 
As we can see from Eq.(\ref{eq1}),  a non-decoupling limit can be realized when   $\Delta$ and $\bar{m}^2$ grow  simultaneously to large values, by keeping the 
$\xi$ ratio  nonvanishing. Under this requirement,  Eq.(\ref{eq1}) shows that the scale $\Lambda_{eff}$ has a non-decoupling behavior, being proportional to the Higgs vev as $\Lambda_{eff}\sim {\cal O}(v/\xi^2)$. It can then potentially lead to  observable effects even in  case of  a heavy messenger sector~\cite{Gabrielli:2014oya}, since in the GRFM typically one has $\xi\sim$ a few tens \%.
Furthermore, we assume that the lightest messenger mass $m_{-}$ satisfies the lower bound $m_{-} \gsim 2$ TeV, in order to avoid a conflict with present collider limits on the direct search of new colored particles. This also guarantees the validity of the low energy approximation in the effective Lagrangian of Eq.(\ref{LH}).

An unsuppressed Higgs-boson coupling to a photon and a dark photon $H\gamma\bar\gamma$ in the Lagrangian in Eq.(\ref{LH}) 
could then provide a privileged way to search for  dark photons  via Higgs production at colliders, and subsequent $H\to \gamma \bar{\gamma}$ decay. 
In this paper, we consider the case of a massless dark photon, that from a phenomenological point of view is anyhow equivalent to a very light dark photon, which escapes detection by a typical collider apparatus.

 A model-independent (parton-level) analysis of  Higgs production via $gg\to H$ at the LHC as a mean for searching for massless dark photons  has been presented in~\cite{Gabrielli:2014oya} for an  LHC c.m. energy of 8 TeV. More recently, an improved study (including  parton-shower effects) with a c.m. energy upgraded to 14~TeV,  has been done in~\cite{Biswas:2016jsh}, where both the gluon-fusion and the vector-boson fusion (VBF)  production mechanisms have been considered. A crucial point is that, 
since the on-shell massless dark photon can be fully decoupled from SM fermions at tree 
level~\cite{Holdom:1985ag},
it is characterised  by a neutrino-like signature in a normal collider detector. 
After its production in collisions, it can then be revealed only by a missing-energy/missing-momentum  measurement. 
For a Higgs boson at rest, the  corresponding signature is quite striking, 
consisting  of  a monochromatic photon with energy $E_{\gamma}=m_H/2$, and similar amount of missing energy, both resonating at the Higgs mass $m_H$. 
By scrutinizing all the relevant reducible and irreducible backgrounds to the corresponding 
$\gamma + \slashed{E}_T+X$ final state, 
 in the gluon-fusion channel,
a $5\sigma$ statistical sensitivity (needed for discovery) is obtained for a branching ratio ${\rm BR}(H\to \gamma\bar{\gamma})\simeq 0.1\%$  at 14 TeV, with an integrated luminosity of $L\simeq 300\ {\rm fb}^{-1}$~\cite{Biswas:2016jsh}.

The effective vertex $H\gamma\bar\gamma$ in Eq.(\ref{LH}) can be complemented
by an effective $HZ\bar\gamma$ coupling to the $Z$ vector boson. Both can give rise to quite distinctive new signatures at future high-energy linear and circular $e^+e^-$ facilities 
(like ILC~\cite{Behnke:2013xla,Baer:2013cma,Fujii:2015jha}, CLIC~\cite{Aicheler:2012bya}, FCC-ee~\cite{FCC-ee}, CEPC~\cite{CEPC}). 
In particular,  the $e^+e^-\to H\bar{\gamma}$ associated production of a Higgs boson and a massless dark photon via a $\gamma/Z$ exchange in the $s$ channel has been analysed in a model independent way at $\sqrt s\simeq 240$ GeV in 
\cite{Biswas:2015sha}. The corresponding signature consists of  a Higgs boson system 
(with the Higgs mainly decaying  into a $b\bar{b}$ pair) recoiling against a {\it massless} invisible system, which remarkably has no irreducible SM background.

In this paper, we consider a different  $e^+e^-$ channel 
involving the $H\gamma\bar\gamma$ coupling.
We study the  $e^+e^-\to H Z$ associated production (which provides the largest 
Higgs-boson sample), with final states 
corresponding to the $H\to \gamma \bar{\gamma}$ decay.
\begin{figure}
\begin{center}
\includegraphics[width = 0.4\textwidth]{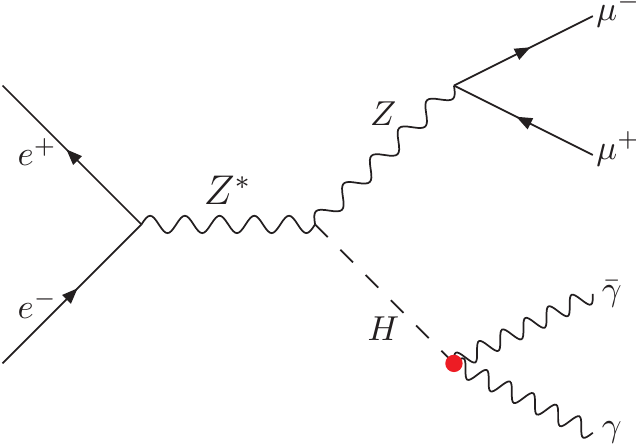}
\hskip 28 pt
\includegraphics[width = 0.4\textwidth]{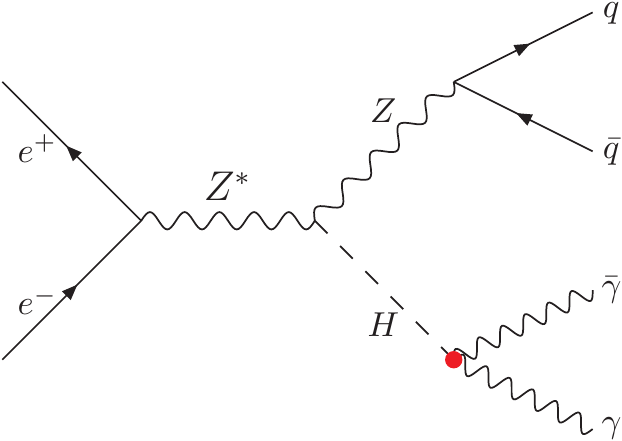}
\caption{Feynman diagrams for $e^+e^- \to ZH \to (\mu^+\mu^-\! ,q\bar{q})(\gamma\bar{\gamma})$. }
\label{FD_ZH}
\end{center}
\end{figure}
 In particular, we will analyse 
both the leptonic  $Z\to \mu^+\mu^-$, and the hadronic  $Z\to q\bar{q}$ decay 
for the $Z$-boson, giving rise, respectively, to the processes
$$
e^+e^-\to ZH \to \mu^+\mu^-\gamma \bar{\gamma}, 
$$
and 
$$
e^+e^-\to ZH \to  q \bar q\, \gamma \bar{\gamma} ,
$$
(depicted in Figure~\ref{FD_ZH}), where, as anticipated, $\bar{\gamma}$ is 
a massless and invisible particle.

The   $\bar \gamma$ production mediated by a Higgs boson 
in $e^+e^-$ collisions can provide  complementary information to the $e^+e^-\to H\bar{\gamma}$ channel.
Just as occurs in the  optimisation of $e^+e^-\to H\bar{\gamma}$ channel, requiring an invisible system with vanishing missing mass in the final state will  help a lot in discriminating the 
$e^+e^-\to ZH \to  Z \gamma \bar{\gamma}$ signal 
 from its backgrounds.
 Comparison with the corresponding ${\rm BR}(H\to \gamma\bar{\gamma})$ experimental  sensitivities from  the study of the $e^+e^-\to H\bar{\gamma}$ channel, and from Higgs production  at the LHC  will be provided, too.

In the following we will start by describing a few features 
of a particular theoretical framework that
can indeed foresee the new decay channel $H\to \gamma \bar{\gamma}$.
On the other hand, we stress that the  results of the present study will be actually model independent. Indeed, 
the phenomenological analysis that will be described will depend by just one new beyond-the-standard-model (BSM) parameter, that is  ${\rm BR}(H\to \gamma\bar{\gamma})$ (assuming that 
possible BSM deviations of other SM couplings entering the amplitude $e^+e^-\to ZH$
are subdominant).

 The paper is organised as follows. In Section~2 we introduce the effective dark-photon couplings to the Higgs boson, and show some relevant model-independent 
 parametrisation
 of the Higgs decay BR's that are affected by the effective couplings. In Section 3 we present the phenomenological analysis of the process  
 $e^+e^-\to ZH \to  Z \gamma \bar{\gamma}$,  we study how to discriminate  the signal and  different backgrounds  for the two final states corresponding to 
  $Z\to \mu^+\mu^-$  and  $Z\to q\bar{q}$, and present the
  corresponding sensitivities in the  ${\rm BR}(H\to \gamma\bar{\gamma})$ measurement. Concluding remarks are given in Section 4.

\section{Theoretical framework}

Here we present the relevant gauge-invariant dark photon effective couplings  to the Higgs boson. Although these couplings will be parametrised in a  model-independent way, we will use the GRFM scenario in~\cite{Gabrielli:2013jka} as a benchmark model which can give rise to these effective interactions.

\begin{figure}
\begin{center}
\includegraphics[width=0.8\textwidth]{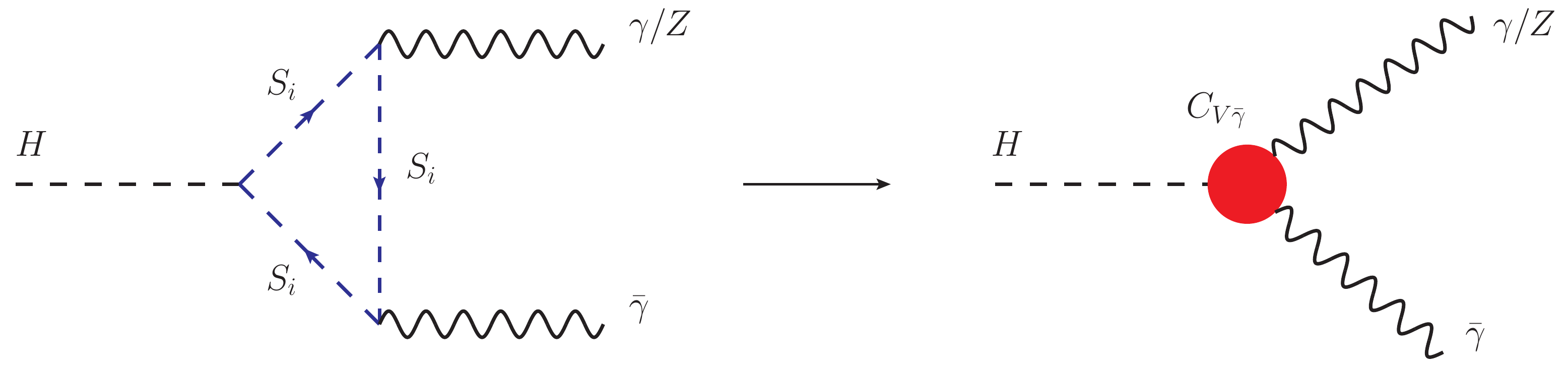}
\caption{Effective coupling approximation for the vertices 
$H \gamma\, \bar{\gamma}$ and $HZ\, \bar{\gamma}$, where $S_i$ are the messenger fields, and in, $C_{V\bar\gamma}$, $V=\gamma,Z$ .}
\label{fig:effective}
\end{center}
\end{figure}

In the GRFM framework, new effective couplings between the 
Higgs, photon and dark photon can be induced at one loop due to the exchange of heavy messenger fields that are  charged under both the SM  and the hidden $U(1)$ gauge groups (Figure~\ref{fig:effective}).
The effective theory approximation can indeed be applied if the messenger sector is much heavier than both the Higgs mass $m_H$ and the dark-fermion masses, as occurs in the GRFM,
 where the condition is automatically satisfied once vacuum stability bounds and 
 dark-matter constraints are applied. In general, 
 the NP sector will also contribute to the  Higgs effective interactions
with two photons, one photon and a $Z$, and two gluons. In the following, we do not consider the latter effects. We anyhow stress that our approach has a more general validity, being applicable to any NP scenario in which there is a heavy messenger sector that couples to both the SM fields and the $U(1)$ dark gauge sector.

In order to provide the formalism for the model independent analysis, we give below the relevant low energy effective Lagrangian 
${\mathcal{L}}_{\rm DP_H}$, connecting the Higgs boson to the dark photon, can be expressed in terms of dimensionless (real) coefficients $C_{ik}$ (with $i,k =\bar \gamma , {\gamma}, Z$) as
\bea
 {\mathcal{L}}_{\rm DP_H} &=& \frac{\alpha}{\pi}\Big(\frac{C_{\gamma \bar{\gamma}}}{v}\gamma^{\mu \nu}\bar{\gamma}_{\mu \nu} H \, +\,  
\frac{C_{Z\bar{\gamma}}}{v} Z^{\mu \nu}\bar{\gamma}_{\mu \nu} H
\, +\, \frac{C_{\bar{\gamma}\bar{\gamma}}}{v} \bar{\gamma}^{\mu \nu}\bar{\gamma}_{\mu \nu} H\Big),
\label{Leff}
\eea
where $\alpha$ is the  SM fine structure constant, $v$ the SM Higgs vacuum expectation value,  and $\gamma_{\mu \nu}$, $Z_{\mu \nu}$, $\bar{\gamma}_{\mu \nu}$ are the  field strentghs of photon, $Z$ boson, and dark photon,  respectively ($\gamma_{\mu \nu}\equiv \partial_{\mu} A_{\nu}-\partial_{\nu} A_{\mu}$ for the photon field $A_{\mu}$).

Following the usual approach, the $C_{ik}$ coefficients  in Eqs.(\ref{Leff}) can be  computed  in the complete theory by evaluating    one-loop amplitudes 
for specific physical processes, and by matching them with the corresponding results obtained at tree level via the effective Lagrangian in Eq.(\ref{Leff}).
The full set of predictions for the $C_{ik}$ coefficients for the GRFM model  can be found in \cite{Biswas:2015sha,Gabrielli:2014oya}.

The basic $C_{ik}$ coefficients in Eq.(\ref{Leff}) can be directly connected to the corresponding Higgs $H\to i \,k$  decay widths.
In particular, for the decay width $\Gamma(H\to \gamma \bar{\gamma} )$, taking into account the parametrization in Eq.(\ref{Leff}), one has
\cite{Gabrielli:2014oya},
\bea
\Gamma(H\to \gamma \bar{\gamma})&=&\frac{m_H^3 \alpha^2 |C_{\gamma\bar{\gamma}}|^2}{8 \pi^3 v^2}\, .
\eea
Analogous results can be obtained for the  $H\to \bar{\gamma}\bar{\gamma}$ and
$H\to Z \bar{\gamma}$ widths by replacing  
$|C_{\gamma \bar{\gamma}}|^2\;$ by $2|C_{\bar{\gamma} \bar{\gamma}}|^2$,
and $|C_{Z \bar{\gamma}}|^2$, respectively.

\begin{figure}
\begin{center}
\includegraphics[width=0.6\textwidth]{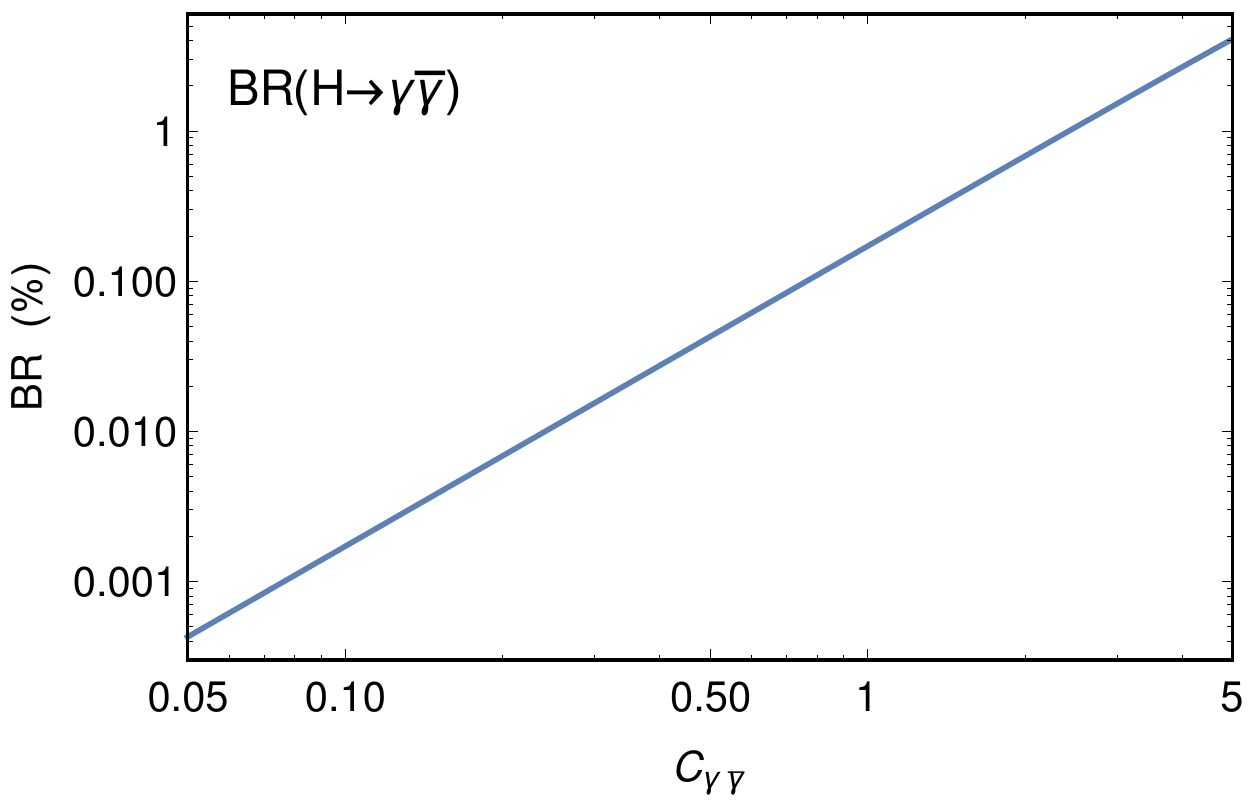}
\caption{Branching ratio for $H\to \gamma \bar{\gamma}$ in percent as a function of the effective coupling  $C_{\gamma\bar{\gamma}}$, for all other effective couplings at their SM values. The $C_{\gamma\bar{\gamma}}$ range in the plot has been choosen such as to cover typical  BR ranges predicted by the GRFM 
  (cf. Figure 1 in \cite{Gabrielli:2014oya}).}
\label{fig:BRversusC}
\end{center}
\end{figure}

In Figure~\ref{fig:BRversusC} we show the  branching ratio for 
$H\to \gamma \bar{\gamma}$ in percent as a function of the corresponding $C_{\gamma\bar{\gamma}}$ coefficient
(when all other effective couplings vanish). The $C_{\gamma\bar{\gamma}}$ range shown in the plot covers values naturally foreseen in the GRFM model.  
One can then  get for the Higgs decays into a dark photon an enhancement 
factor  ${\cal O}(10)$ with respect  to the SM Higgs decays where the dark  photon is replaced by a photon. This makes the corresponding  phenomenology quite relevant for both LHC and future-collider studies.

Neglecting the $C_{Z\bar{\gamma}}$ contribution, 
a convenient model-independent \mbox{BR($H\to \gamma \bar{\gamma},\; \bar \gamma \bar{\gamma}, \;  \gamma {\gamma}  $)} 
parametrisation can be provided, involving the 
{\it relative} exotic contributions $r_{ik}$ to the  $H\to i \,k$ decay widths, with  $i,k= \gamma , \bar{\gamma}$, 
where the $r_{ik}$ ratios are defined as
\bea
r_{ik}&\equiv&\frac{\Gamma^{\rm NP}_{ik}}{\Gamma^{\rm SM}_{\gamma\gamma}}\; ,
\label{rij}
\eea
and $\Gamma^{\rm NP}_{i k}$ stands for the pure NP contribution to the $H\to  i\,k$ decay width\footnote{Note that in case of $\Gamma^{\rm NP}_{\gamma\gamma}$, this quantity  is connected to a physical decay width only up to possible interference terms between the SM and the NP $H\to \gamma\gamma$ amplitudes.}. 
Then, the following  model-independent parametrisation of the  quantities 
{BR$_{\gamma\bar{\gamma},\; \bar \gamma \bar{\gamma}, \;  \gamma {\gamma}}
\equiv\, $BR$(H\to {\gamma}\bar{\gamma},\; \bar \gamma \bar{\gamma}, \;  \gamma {\gamma} )$} as functions of $ r_{ik}$ holds \cite{Gabrielli:2014oya} 
\bea
B\!R_{\gamma\bar{\gamma}}&=&B\!R^{\rm SM}_{\gamma\gamma}
\frac{r_{\gamma\bar{\gamma}}}
{1+r_{\bar{\gamma}\bar{\gamma}}B\!R^{\rm SM}_{\gamma\gamma}}\, ,
\nonumber\\
B\!R_{\bar{\gamma}\bar{\gamma}}&=&B\!R^{\rm SM}_{\gamma\gamma}
\frac{r_{\bar{\gamma}\bar{\gamma}}}
{1+r_{\bar{\gamma}\bar{\gamma}}B\!R^{\rm SM}_{\gamma\gamma}}\, ,
\nonumber\\
B\!R_{\gamma\gamma}&=&B\!R^{\rm SM}_{\gamma\gamma}
\frac{\left(1+\chi \sqrt{r_{\gamma\gamma}}\right)^2}
{1+r_{\bar{\gamma}\bar{\gamma}}B\!R^{\rm SM}_{\gamma\gamma}}\, ,
\label{BRS}
\eea
where  $\chi=\pm 1$ parametrises the relative sign between the SM and the NP loop amplitudes. 

We stress that, in any model where the effective couplings in Eq.~(\ref{Leff}) are generated radiatively by charged messenger fields circulating in the loop, the factors $r_{ik}$ (where $i,k=\gamma,\bar{\gamma},Z$) are not independent, but are determined by the hypercharge assignment of the mediators, as described in \cite{Biswas:2015sha}. 

A consequence of Eq.~(\ref{BRS}) is that these scenarios can also be indirectly constrained by a precision measurement of the Higgs branching ratios for the more-standard decays into two photons or invisible final states.

\section{Collider Analysis}

In this section we discuss the experimental strategies relevant to make a 
measurement of  BR$_{\gamma\bar{\gamma}}$, the Higgs decay BR 
into a photon and an invisible massless dark photon,   via 
 the process $e^+e^-\rightarrow ZH$ followed by  $H\rightarrow \gamma\bar{\gamma}$
 in an $e^+e^-$ collider with  cm energy of about 240 GeV, which maximises the Higgs cross section. 
 This setup could be realised at either linear (like ILC) or circular (like FCC-ee
 and  CEPC)
 facilities  with integrated luminosities up to about 10 ab$^{-1}$ at 240 GeV, corresponding to the production of up to about 2 million  Higgs bosons.

 We outline the search strategies for both the leptonic $Z\rightarrow \ell^+\ell^-$ and hadronic $Z\rightarrow q\bar{q}$ final states (cf. Figure \ref{FD_ZH}).  Being stable and escaping the detection, a massless dark photon shows up in normal detectors like a neutrino.
 Thus the $e^+e^-\rightarrow ZH$ leptonic final state consists of a pair of opposite-sign same-flavor leptons, a photon, and missing energy/momentum (named $\slashed E/\slashed p$), whereas the hadronic final state contains two jets, a photon, and missing energy/momentum.

We have simulated the signal and SM backgrounds with MadGraph5\_aMC@NLO \cite{MG5_aMC@NLO} interfaced with PYTHIA \cite{pythia6} 
to include the initial and final state radiation and hadronisation effects\footnote{Initial state radiation effects considered here will be typical of circular $e^+e^-$ colliders,
as we will disregard possible beamstrahlung effects.}.
The jets are clustered using a simple cone algorithm 
with cone size $R=0.4$ and transverse momentum $p_T>20$ GeV.

We assume the following specification for the detector performance 
\cite{Cerri:2016bew,Behnke:2013lya}: 

\begin{itemize}
\item Muon momentum resolution: $\Delta p/p = 0.1\% + p_T/(10^5\ {\rm GeV})$ for $|\eta|<1$, and 10 times poorer  for $1<|\eta|<2.5$.
\item Photon energy resolution: $\Delta E/E = 16.6\%/\sqrt{E/\ {\rm GeV}}+1.1\%$.
\item Jet energy resolution: $\Delta E/E = 30\%/\sqrt{E/\ {\rm GeV}}$
\item Particle identification efficiency for muons and photons: $99\%$ for $p_T>10\ {\rm GeV}$.
\end{itemize}

\subsection{Leptonic channel: $e^+e^-\rightarrow ZH\rightarrow \mu^+\mu^- \gamma\bar{\gamma}$}
Thanks to the superior momentum resolution, the leptonic channel is the cleanest of the final states, as the leptonic $Z$ can be reconstructed very efficiently. Since the muon momentum resolution is better than the one for electrons, we outline here the search for the $Z\rightarrow \mu^+\mu^-$ channel. The electron channel will contribute less 
to the total $e^+e^-\rightarrow ZH$ sensitivity not only for the poorer electron momentum resolution, but also for the additional SM neutral-current $t$-channel 
$e^+e^-\rightarrow  e^+e^- \bar \nu {\nu} \gamma$
component in the background, which has no equivalent for the muonic final state. 
Initially, we select the events containing two opposite-sign muons and a single photon with the following {\it basic cuts}:
\begin{itemize}
\item muon and photon transverse momentum with $\,p_T^\mu, p_T^\gamma > 10\ {\rm GeV}$,
\item muon and photon pseudorapidity in the range $|\eta^\mu|, |\eta^\gamma| < 2.5$,
\item missing energy with $\slashed{E} > 10\ {\rm GeV}$.
\item angular separation between any two objects with \mbox{$\Delta R > 0.2$},
\item jet veto for $p_T^j>20\ {\rm GeV}$.
\end{itemize}

The irreducible SM background for the $e^+e^-\rightarrow ZH\rightarrow \!\mu^+\mu^- \gamma\bar{\gamma}$ final state is given by  the process $e^+e^-\rightarrow \!\mu^+\mu^-\nu\bar{\nu}\gamma$, which arises from the resonant contribution of the channels $e^+e^-\rightarrow ZZ\gamma$ and $e^+e^-\rightarrow WW\gamma$, as well as from different  $t$-channel processes such as $e^+e^-\rightarrow\nu\bar{\nu}Z\gamma$. 
In the analysis of the irreducible $\mu^+\mu^-\nu\bar{\nu}\gamma$ background 
both the individual resonant $WW\gamma$ and $ZZ\gamma$ components will be analysed 
in parallel to the inclusive $\mu^+\mu^-\!\nu\bar{\nu}\gamma$ production.
Then, there are reducible backgrounds from $Z\gamma$ events accompanied by fake missing energy, which can originate from initial state radiation/beamstrahlung, mismeasurement of the lepton or photon momenta, or missed final-state objects. The last category contains 
 the $e^+e^- \rightarrow ZH \rightarrow \mu^+\mu^-\gamma\gamma$ process when one of the photons escapes detection. 
The latter events will have the same kinematic features as
 the signal, but  rates  suppressed by both  BR($H\rightarrow\gamma\gamma)\simeq 2\times 10^{-3}$ and   the small probability of missing one of the photons while the other goes inside the central barrel and passes the event selection. Further details will follow on the (in general negligible) $H\to \gamma\gamma$   contribution to the background\footnote{ We have also scrutinized the nonresonant $e^+e^-  \rightarrow \mu^+\mu^-\gamma\gamma$ channel, and found that in general 
 this background can be controlled  by demandingÊ
an extra missing transverse-energy lower cut of a few GeV's over the final cut flow, without affecting our presentÊ
analysis.}.

\begin{figure}
\begin{center}
\includegraphics[width = 0.49\textwidth]{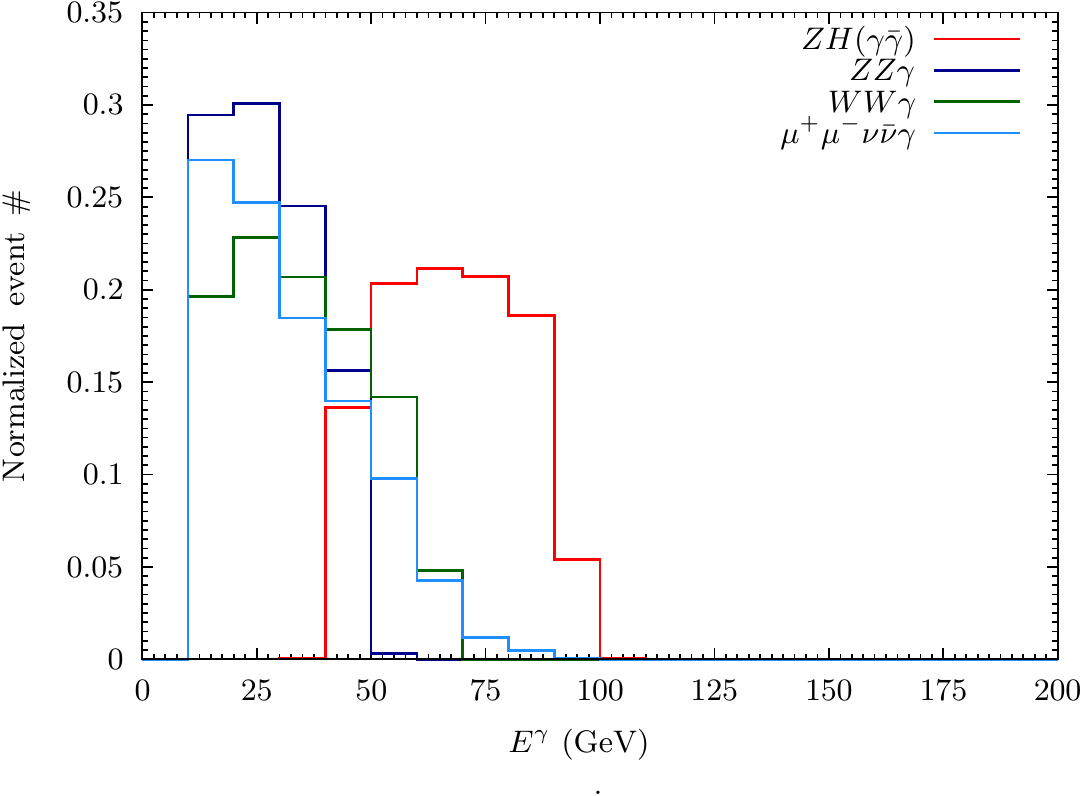}
\includegraphics[width = 0.49\textwidth]{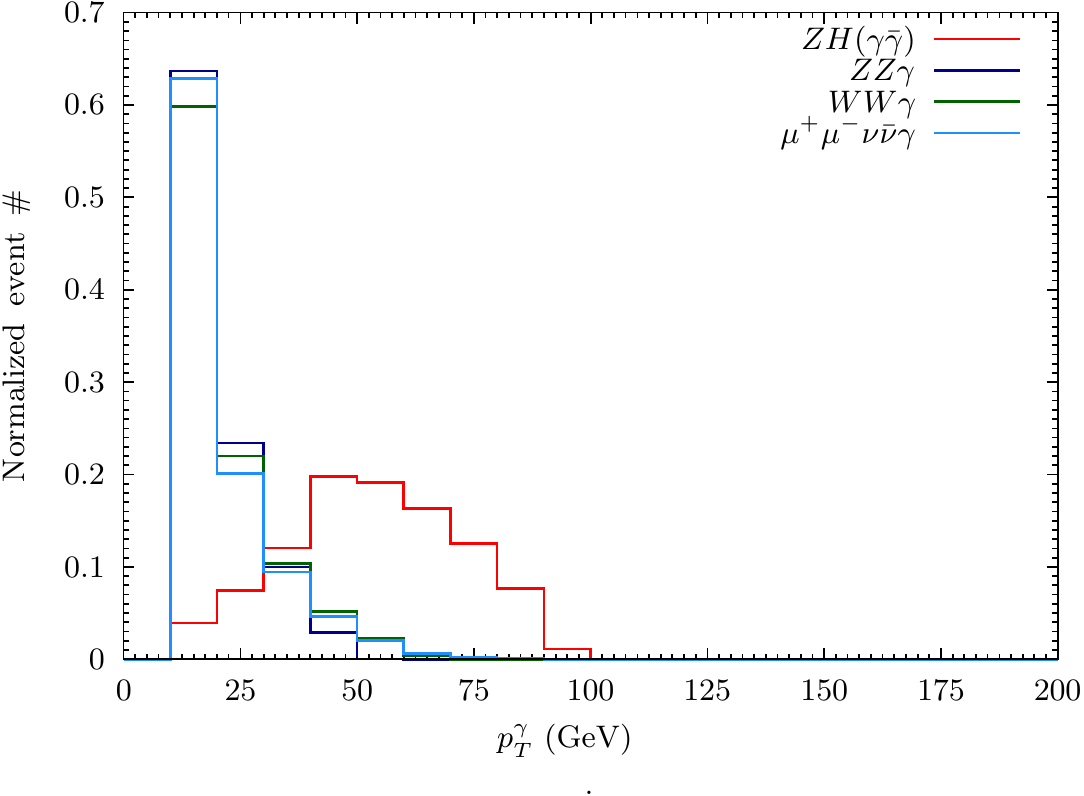}
\caption{The photon energy and transverse momentum distributions  for the 
$e^+e^-\!\!\rightarrow\mu^+\mu^-\gamma\bar\gamma$ signal  and $e^+e^-\!\!\rightarrow \!\mu^+\mu^-\nu\bar{\nu}\gamma$ background,  after applying the set of basic  cuts, at $\sqrt s=240$~GeV. Results for the individual resonant  $WW\gamma$ and $ZZ\gamma$ background components are also shown.}
\label{muon channel photon pt-pe}
\end{center}
\end{figure}
The photon energy and transverse momentum normalised distributions are shown
in Figure~\ref{muon channel photon pt-pe} 
both for signal and main backgrounds, after implementing the above list of basic cuts.

Apart from the latter distributions, signal events can be particularly discriminated by the use of a few kinematic variables characterising them. Three variables are of special interest: the missing mass $M_{\rm miss}$, the invariant mass of the photon-missing-energy system $M_{\gamma\bar{\gamma}}$, and the invariant mass of the lepton pair $M_{\ell\ell}$. These are defined as
\begin{eqnarray}
M_{\rm miss} &=& \sqrt{\slashed{E}^2-\slashed{\vec{p}}^2}, \\
M_{\gamma\bar{\gamma}} &=& \sqrt{2(E_\gamma\slashed{E}-\vec{p}_\gamma\cdot\slashed{\vec{p}} ) }, \\
M_{\ell\ell} &=& \sqrt{2(E_{\ell^+}E_{\ell^-}-\vec{p}_{\ell^+}\cdot \vec{p}_{\ell^-})},
\end{eqnarray}
where the missing energy $\slashed{E}$ and momentum $\slashed{\vec{p}}$ are experimentally defined by the equations $\slashed{E} = \sqrt{s}-\sum_i E_i$ and $\slashed{\vec{p}} = -\sum_i\vec{p}_i$ (the sum is over all detected final particles). For the signal events, where the missing energy is carried by the massless dark photon, these variables are centered at $M_{\rm miss} = 0$, $M_{\gamma\bar{\gamma}} = m_H$ and $M_{\ell\ell} = M_Z$.

\begin{figure}
\begin{center}
\includegraphics[width = 0.49\textwidth]{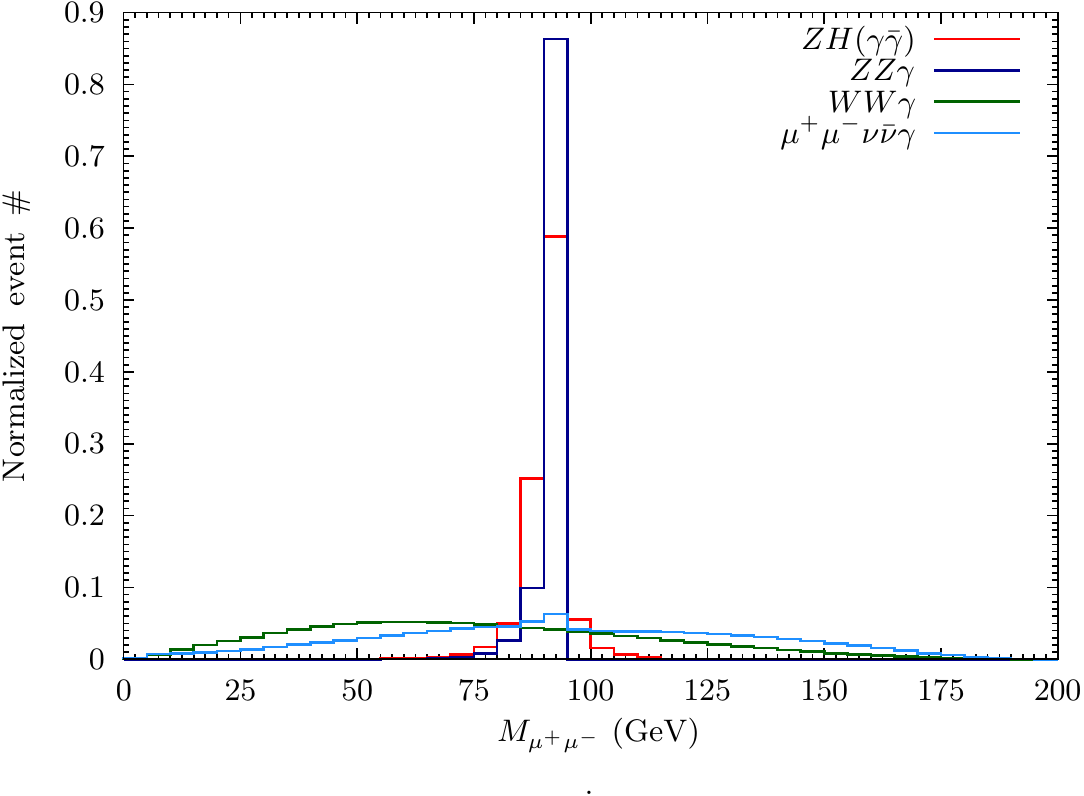}
\includegraphics[width = 0.49\textwidth]{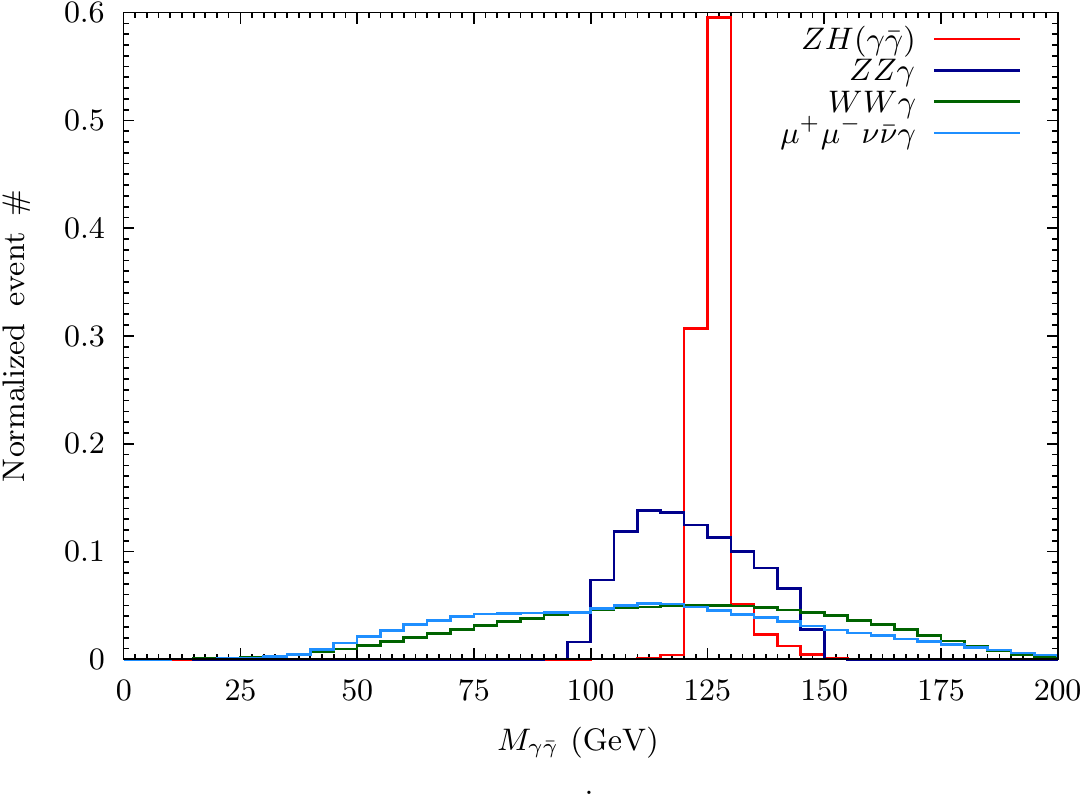}
\caption{The ${\mu^+\mu^-}$ and ${\gamma\bar{\gamma}}$ invariant-mass distributions 
for the $e^+e^-\!\!\rightarrow\mu^+\mu^-\gamma\bar\gamma$ signal  and $e^+e^-\!\!\rightarrow \!\mu^+\mu^-\nu\bar{\nu}\gamma$ background, for $\sqrt s=240$~GeV. The $M_{\mu^+\mu^-}$ distributions is obtained after imposing just 
the set of basic cuts described in the text, whereas the $M_{\gamma\bar{\gamma}}$ distribution is affected
by an additional cut \mbox{$86\ {\rm GeV} < M_{\mu^+\mu^-} < 96\ {\rm GeV}$}.
Results for the individual resonant  $WW\gamma$ and $ZZ\gamma$ background components are also shown.}
\label{muon channel Z and H invmass}
\end{center}
\end{figure}
The $M_{\mu^+\mu^-}$ and $M_{\gamma\bar{\gamma}}$ normalised distributions for the signal and  SM-background events are shown in Figure~\ref{muon channel Z and H invmass}. 
The $M_{\mu^+\mu^-}$ distribution is obtained assuming the basic cuts listed above. An additional cut $86\ {\rm GeV} < M_{\mu^+\mu^-} < 96\ {\rm GeV}$ has been applied before plotting the $M_{\gamma\bar{\gamma}}$ distribution.

\begin{figure}
\begin{center}
\includegraphics[width = 0.49\textwidth]{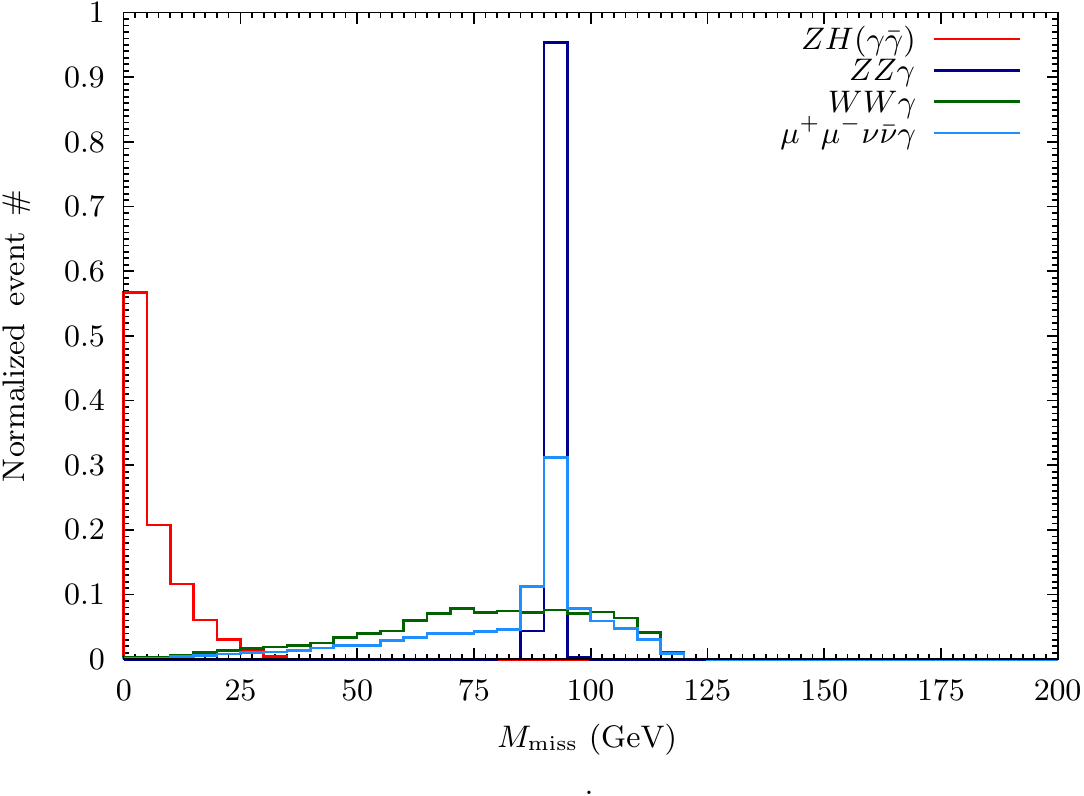}
\includegraphics[width = 0.49\textwidth]{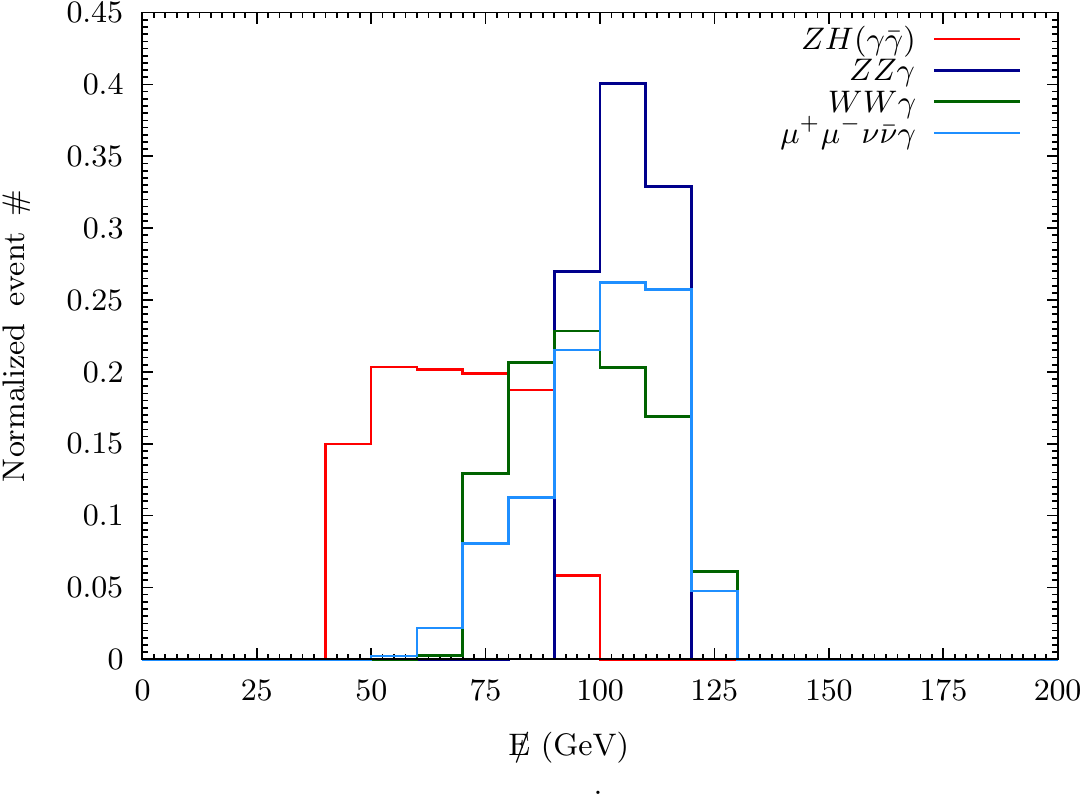}
\caption{The missing-mass  and missing-energy distributions 
for the $e^+e^-\!\!\rightarrow\mu^+\mu^-\gamma\bar\gamma$ signal  and $e^+e^-\!\!\rightarrow \!\mu^+\mu^-\nu\bar{\nu}\gamma$ background, for $\sqrt s=240$~GeV, after imposing the 
invariant mass cuts around the $M_Z$ and $m_H$ on the ${\mu^+\mu^-}$ and ${\gamma\bar{\gamma}}$ systems, respectively. }
\label{muon channel  M_miss and E_miss}
\end{center}
\end{figure}

We therefore suppress the SM background   by  the following selection criteria imposed on top of the basic cuts:

\begin{itemize}
\item $Z$ mass cut: $86\ {\rm GeV} < M_{\mu^+\mu^-} < 96\ {\rm GeV}$,
\item Higgs mass cut: $120\ {\rm GeV} < M_{\gamma\bar{\gamma}} < 130\ {\rm GeV}$.
\end{itemize}
After applying the above two cuts, one obtains the $M_{\rm miss}$ and $\slashed{E}$ normalised distributions  shown 
in Figure~\ref{muon channel  M_miss and E_miss}.
Because of the signal low-mass structure in the $M_{\rm miss}$ distribution in 
 Figure~\ref{muon channel  M_miss and E_miss}, we then impose the additional cut
\begin{itemize}
\item Missing mass cut: $M_{\rm miss} < 20\ {\rm GeV}$.
\end{itemize}

Cutting away large $M_{\rm miss}$ values 
proves indeed very effective for background suppression, since most of the background sub-processes contain massive invisible systems  which are not likely to have low 
$M_{\rm miss}$.

We then stop our cut flow, 
since, after applying the $M_{\rm miss}$ optimisation on distributions in Figures~\ref{muon channel  M_miss and E_miss},  the $\slashed{E}$ distribution  (that is largely correlated to the 
$M_{\rm miss}$ distribution) does not offer extra handle for further optimization.

We now comment on  the reducible SM contribution to the background coming from 
$e^+e^-\to ZH \to \mu^+\mu\!^-\gamma\gamma$, where one of the photons in the 
$H\to \gamma\gamma$ decay is not identified. Indeed, 
some $\slashed{E}$ can come from  either energy mismeasurement or the unlikely situation where just one of the 
photons lies  in the forward region ($|\eta|>5$) and is not detected, or a combination 
of both.  For $BR_{\gamma\bar{\gamma}}=1\%$, we checked  that the $ZH\to Z\gamma\gamma$ 
background is suppressed by two order of magnitudes with respect to the signal 
 (by imposing the cut flow in table~\ref{muon channel cut flow}).
 For $BR_{\gamma\bar{\gamma}} \simeq BR_{\gamma{\gamma}}$, 
 the number of signal events is still about 30 times the number of this  background events.  

\begin{figure}
\begin{center}
\includegraphics[width = 0.7\textwidth]{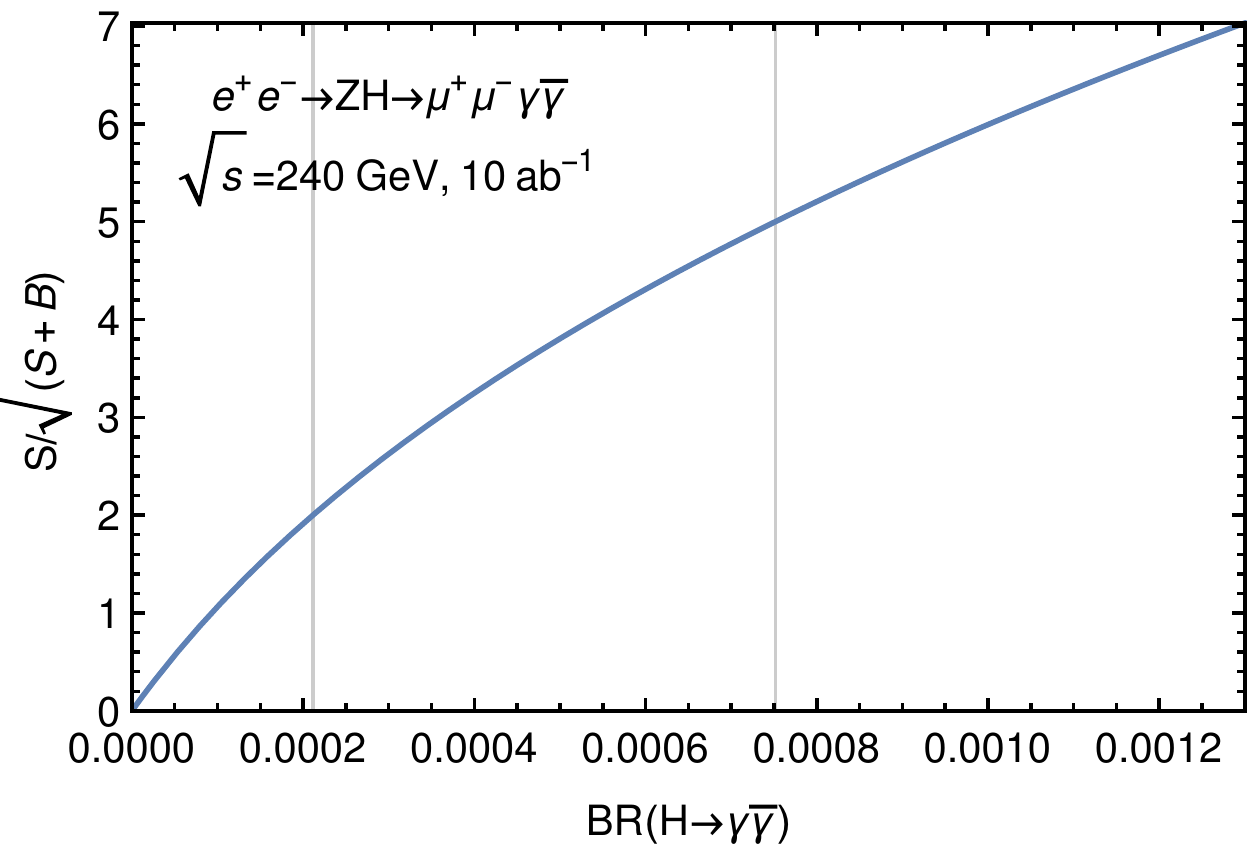}
\caption{Signal significance for the $e^+e^-\!\!\rightarrow ZH\rightarrow \mu^+\mu^- \gamma\bar{\gamma}$ channel  versus $BR_{\gamma\bar{\gamma}}$ for \mbox{10 ab$^{-1}$} at 240 GeV. The left vertical grey line corresponds to a 95\% CL exclusion, while the right line
points to the 5$\sigma$ discovery reach.}
\label{muon channel significance}
\end{center}
\end{figure}

The effect of these cuts on the signal and inclusive background event yields is presented in 
table~\ref{muon channel cut flow}. The resulting significance $S/\sqrt{S+B}$ (where $S$ is the number of signal events and $B$ the number of background events) is shown as a function of  $BR_{\gamma\bar{\gamma}}$ in Figure~\ref{muon channel significance}, assuming an integrated luminosity of $10\ {\rm ab}^{-1}$ at  $\sqrt{s} = 240\ {\rm GeV}$. We find that in the leptonic channel one can exclude values down to $BR_{\gamma\bar{\gamma}}=2\times 10^{-4}$ 
at 95\% C.L., while the  $5\sigma$ discovery reach is  
$BR_{\gamma\bar{\gamma}}=7.5\times 10^{-4}$.
\begin{table}
\begin{center}
\begin{tabular}{|l||c|c|c|c|}
\hline
Process & Basic cuts  & $M_{\ell\ell}$ cut & $M_{\gamma\bar{\gamma}}$ cut & $M_{\rm miss}$ cut \\
\hline
$\mu^+\mu^- \gamma\bar{\gamma}$ \;\;($BR_{\gamma\bar{\gamma}} = 0.1\%$) & 65.3 & 54.9 & 49.7 & 47.3 \\
\hline
$\mu^+\mu^-\nu\bar{\nu}\gamma\;\;$  & $5.00\times 10^{4} $& $5.73 \times 10^{3}$& $1.09 \times 10^{3}$& 15 \\
\hline
\end{tabular}
\caption{Event yields after sequential cuts for $e^+e^-\!\!\rightarrow ZH\rightarrow \mu^+\mu^- \gamma\bar{\gamma}$ and corresponding background, for an integrated luminosity of $10\ {\rm ab}^{-1}$, and  c.m. energy  $\sqrt{s} = 240\ {\rm GeV}$. The signal yield has been normalised assuming $BR_{\gamma\bar{\gamma}}=0.1\%$.} 
\label{muon channel cut flow}
\end{center}
\end{table}      

\subsection{Hadronic channel: $e^+e^-\rightarrow ZH\rightarrow q\bar q  \gamma\bar{\gamma}$}

The worse energy resolution for jets with respect to muons, resulting in a less clean reconstruction of the hadronic   $Z$-boson decay, can be  compensated by the larger $Z$ branching ratio into jets, and the increased phase-space acceptance for jets. It is then important to include 
 the $Z$ hadronic decay mode in the present analysis.

The $e^+e^-\!\!\rightarrow ZH\rightarrow q\bar q  \gamma\bar{\gamma}$ signal consists of two jets, a single photon, and missing energy. 
The main irreducible SM background comes from the process $e^+e^-\rightarrow q\bar q\nu\bar{\nu}\gamma$, which, as we will show in the following, can be effectively suppresed 
by imposing an upper missing-mass cut. 
The main reducible and dominant background arises instead from the jet-pair production accompanied by a hard photon, $e^+e^-\rightarrow q\bar q\gamma\to jj\gamma$. Here, some missing energy is generated either from jet-energy mismeasurement, or, more importantly, by neutrinos generated by  heavy-flavor decays inside the jet showering. The $jj\gamma$ background is then characterised by relatively low values of missing energy and by the approximate alignment of the missing momentum  with one of the jets. 

We perform the initial event selection according to the following {\it basic cuts}:
\begin{itemize}
\item lepton veto for $p_T^\ell >10\ {\rm GeV}$ and $|\eta^\ell|<2.5$,
\item for the photon transverse momentum and pseudorapidity: $p_T^\gamma > 10$ GeV, $|\eta^\gamma|<2.5$,
\item for the jet transverse momentum and pseudorapidity: $p_T^j > 20$ GeV, $|\eta^j|<5.0$,
\item for the missing energy: $\slashed{E} > 10\ {\rm GeV}$.
\item for the angular separation between any pair of visible objects: $\Delta R > 0.4$.
\end{itemize}

We use the same kinematical variables adopted in the lepton-channel analysis, with the obvious replacement of $M_{\ell\ell}$ with the jet-pair invariant mass $M_{jj}$. 

Then, for the signal events, where the missing energy is carried by the massless dark photon, the relevant variables are centered at $M_{\rm miss} = 0$, $M_{\gamma\bar{\gamma}} = m_H$, and $M_{jj} = M_Z$.

\begin{figure}
\begin{center}
\includegraphics[width = 0.49\textwidth]{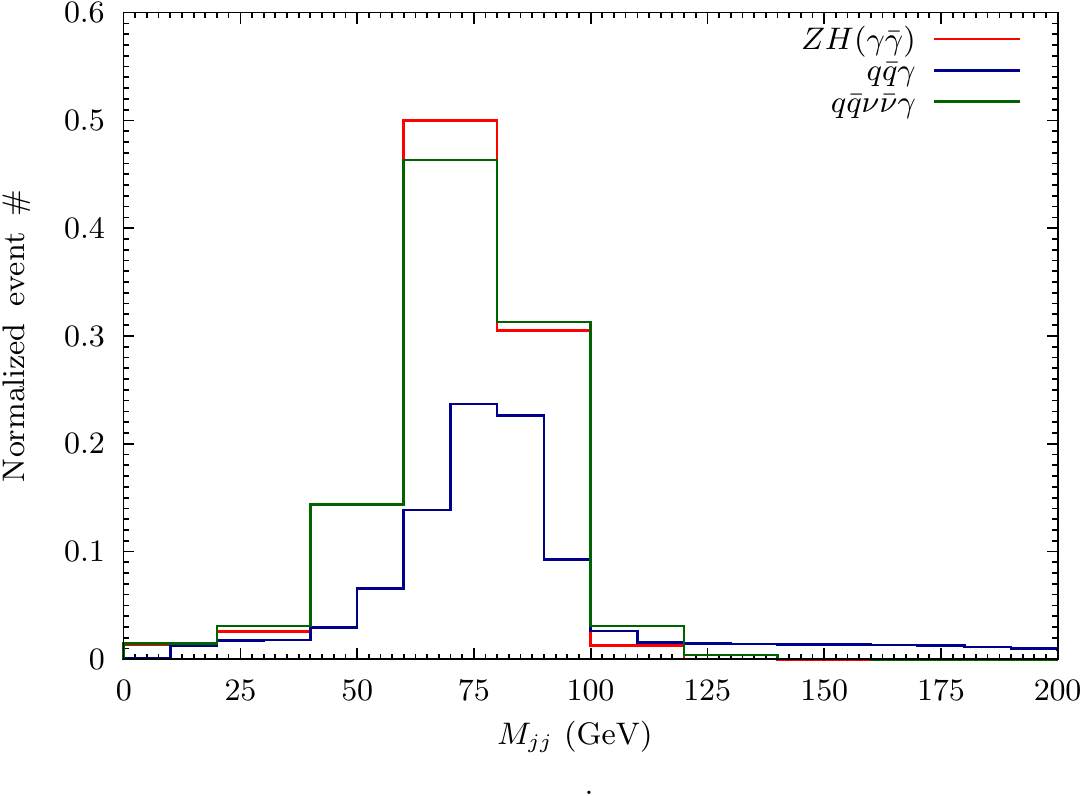}
\includegraphics[width = 0.49\textwidth]{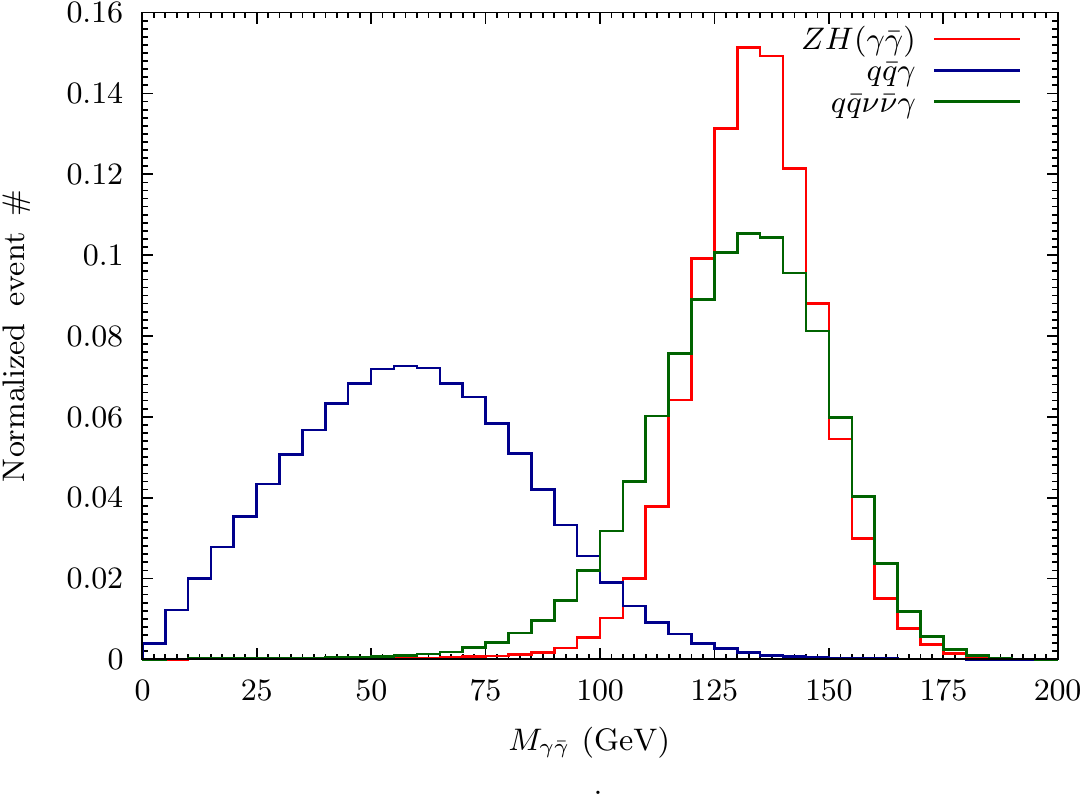}
\caption{The $jj$ and ${\gamma\bar{\gamma}}$ invariant mass distributions 
for the $e^+e^-\!\!\rightarrow ZH\!\rightarrow \!q\bar q  \gamma\bar{\gamma}$ signal  and backgrounds, for $\sqrt s=240$~GeV. The $M_{jj}$ distribution is obtained after imposing
the set of basic cuts described in the text, whereas the $M_{\gamma\bar{\gamma}}$ distribution is obtained
with an additional  \mbox{$50\ {\rm GeV} < M_{jj} < 90\ {\rm GeV}$} cut.}
\label{hadron channel Z and H invmass}
\end{center}
\end{figure}

The $M_{jj}$ and $M_{\gamma\bar{\gamma}}$ normalised distributions for the signal and  SM-background events are shown in Figure \ref{hadron channel Z and H invmass}. 
The $M_{jj}$ distribution is obtained assuming the basic cuts listed above. 
An additional cut  \mbox{$50\ {\rm GeV} < M_{jj} < 90\ {\rm GeV}$} has been applied before plotting the $M_{\gamma\bar{\gamma}}$ distribution
(due to the relatively poor jet-energy resolution, the $M_{jj}$ cut  around the $Z$-boson mass is looser than the $M_{\mu^+\mu^-}$ cut for the leptonic channel).

In Figure~\ref{hadron channel Z and H invmass}, one can see how the extra 
missing-momentum system arising from the $Z\to \bar q q$ showering widens up 
the signal $M_{\gamma\bar{\gamma}}$ peak structure around $m_H$  
with respect to the leptonic-channel $M_{\gamma\bar{\gamma}}$ distribution 
in Figure~\ref{muon channel Z and H invmass}.
Nevertheless, we found that loosening the $120\ {\rm GeV} < M_{\gamma\bar{\gamma}} < 130\ {\rm GeV}$ cut (applied in the leptonic channel) in order to increase the signal statistics induces a milder kinematical characterisation
of the signal events, contaminating them with extra missing energy not originating from the dark photon. This in turn would make further cuts on the $M_{miss}$ less effective for separating the signal from the $q\bar q\gamma$ background.

\begin{figure}
\begin{center}
\includegraphics[width = 0.49\textwidth]{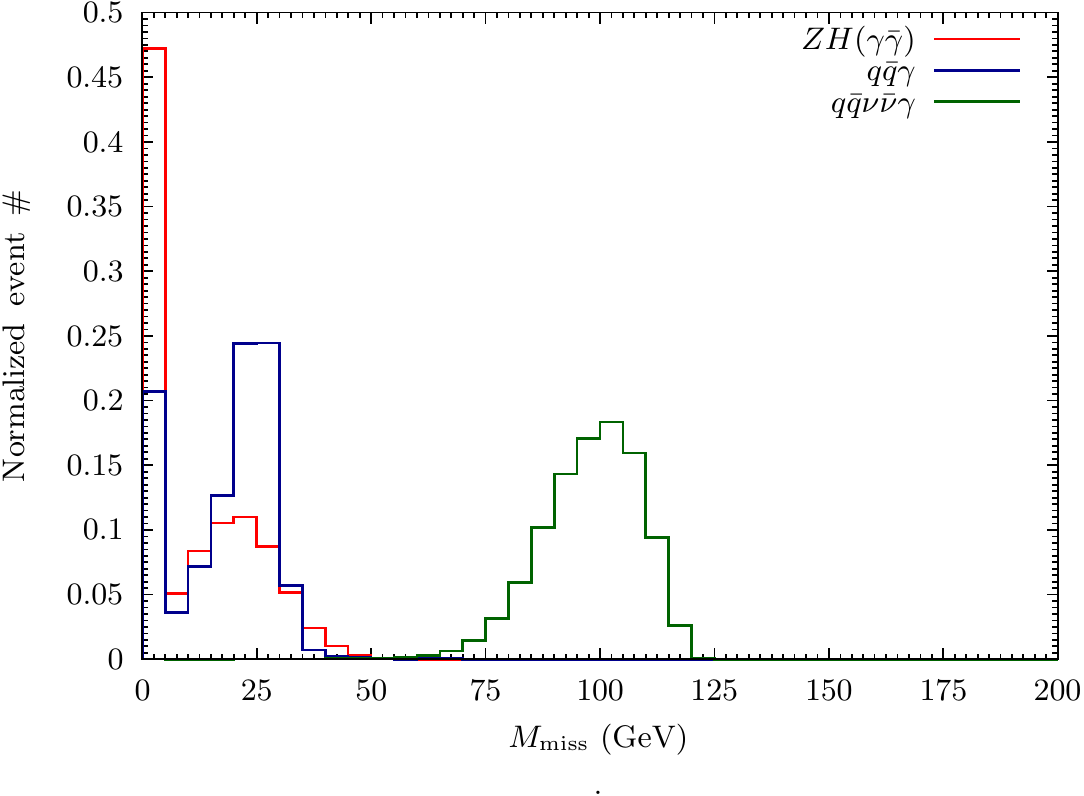}
\includegraphics[width = 0.49\textwidth]{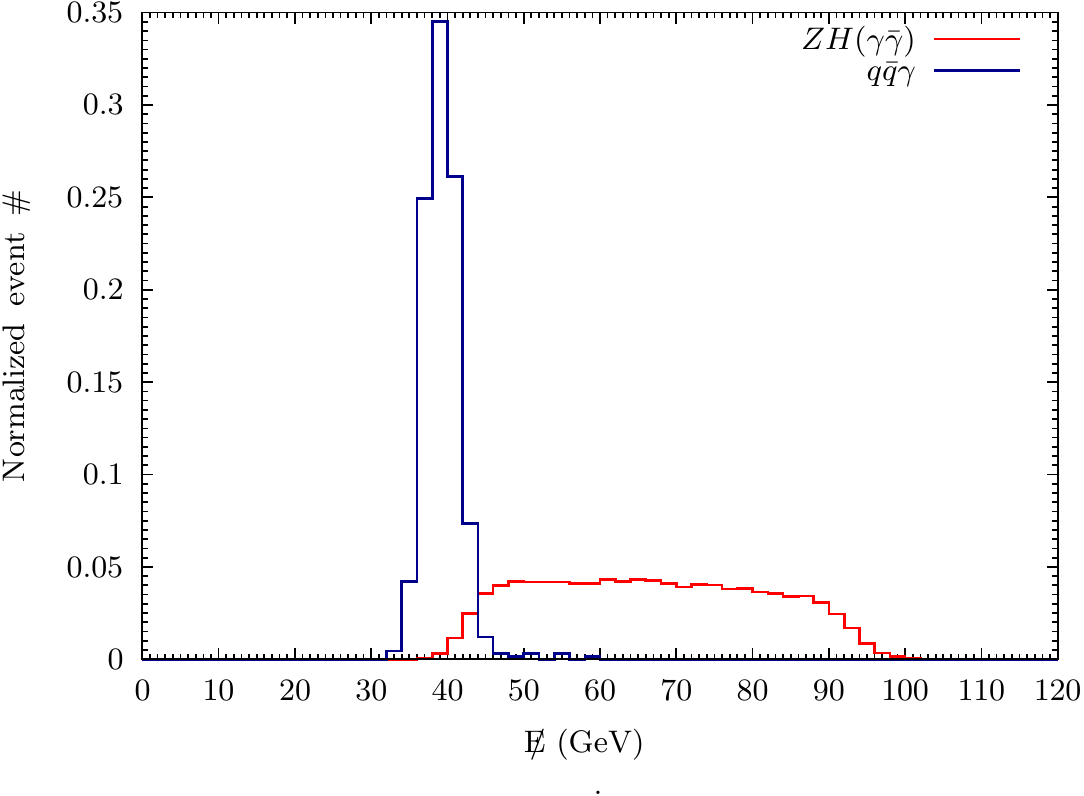}
\caption{The missing mass and missing energy distributions 
for the $e^+e^-\!\!\rightarrow ZH\!\rightarrow \!q\bar q  \gamma\bar{\gamma}$ signal  and corresponding backgrounds,
for $\sqrt s=240$~GeV. The $M_{\rm miss}$ distribution is obtained after imposing   
invariant mass cuts  on the $jj$ and ${\gamma\bar{\gamma}}$  systems around $M_Z$ and $m_H$, respectively, as described in the text. In the $\slashed{E}$ distributions, an additional 
$M_{\rm miss} < 20\ {\rm GeV}$ cut is imposed.}
\label{hadron channel M_miss and E_miss}
\end{center}
\end{figure}

As a consequence, we stick to the {\it narrow} $120\ {\rm GeV} < M_{\gamma\bar{\gamma}} < 130\ {\rm GeV}$ cut, hence selecting signal events where the missing momentum is 
mostly associated to the dark photon. This 
 is anyhow very effective in reducing the $q\bar q\gamma$ background (cf. Figure \ref{hadron channel Z and H invmass}). 
 After that, one obtains the $M_{\rm miss}$  normalised distribution  shown in 
 Figure~\ref{hadron channel M_miss and E_miss} (left plot). Hence, 
  requiring $M_{\rm miss} < 20\ {\rm GeV}$ effectively kills the irreducible 
$q\bar q\nu\bar{\nu}\gamma$ background, 
with a more moderate effect on the $q\bar q\gamma$ reducible component. 
 
In Figure~\ref{hadron channel M_miss and E_miss} (right plot), we have  imposed an additional $M_{\rm miss} < 20\ {\rm GeV}$ cut on the normalised 
 $\slashed{E}$ distribution. In order 
to further mitigate the remaining $q\bar q\gamma$ background, 
one can cut away the region $\slashed{E} \lsim 50\ {\rm GeV}$. We then add a further optimised missing-energy cut   $\slashed{E} > 59\ {\rm GeV}$ to the cut flow.  After that also the $q\bar q\gamma$ background is reduced to a negligible level, and the search, assuming a reference decay rate $BR_{\gamma\bar{\gamma}}=0.1\%$, becomes essentially a counting experiment for the signal events.

\begin{figure}
\begin{center}
\includegraphics[width = 0.7\textwidth]{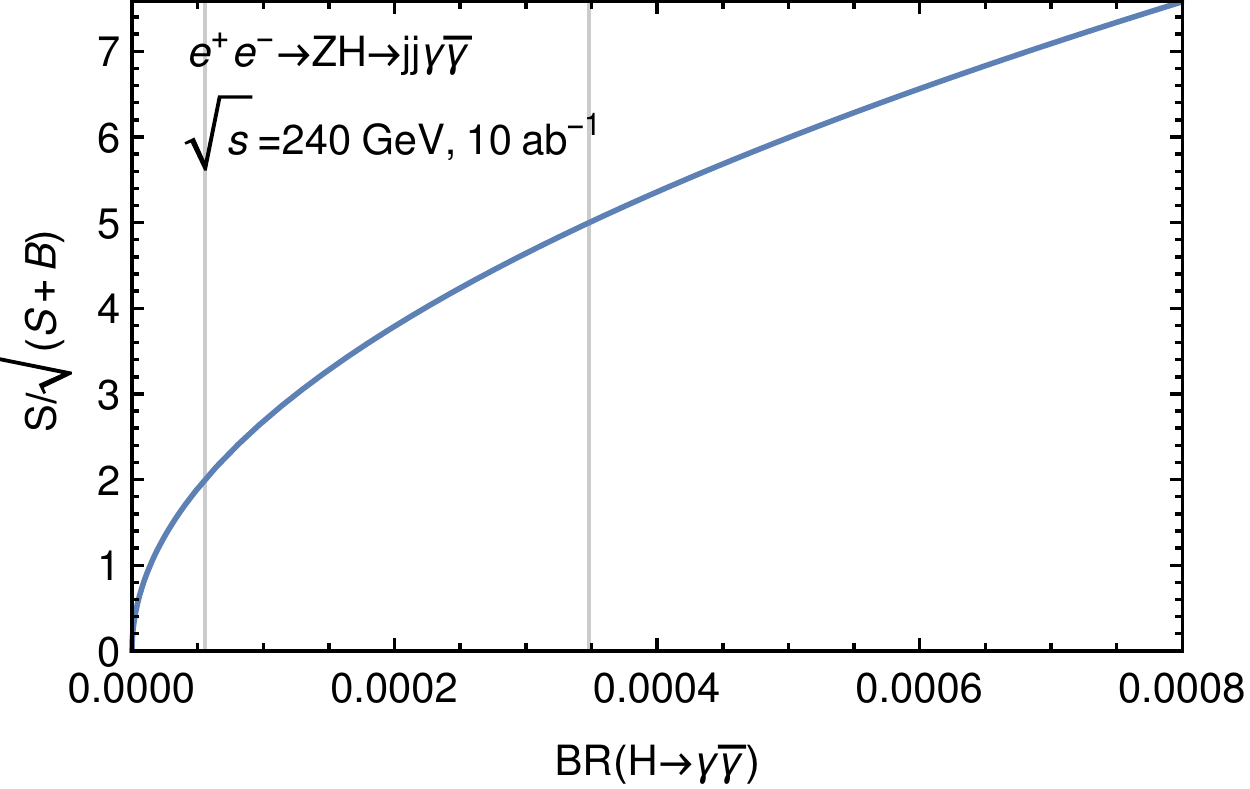}
\caption{Signal significance for the $e^+e^-\!\!\rightarrow ZH\rightarrow \!q\bar q \gamma\bar{\gamma}$ channel  versus $BR_{\gamma\bar{\gamma}}$ for \mbox{10 ab$^{-1}$} at 240 GeV. The left vertical grey line corresponds to a 95\% CL exclusion, while the right line
points to the 5$\sigma$ discovery reach.}
\label{hadron channel significance}
\end{center}
\end{figure}

The effect of the cut flow on the event yields for the signal (for $BR_{\gamma\bar{\gamma}} = 0.1\%$), and backgrounds is shown in 
table~\ref{hadron channel cut flow}, assuming an integrated luminosity of $10\ {\rm ab}^{-1}$. 
 In 
Figure~\ref{hadron channel significance}, the resulting significance is shown as a function of $BR_{\gamma\bar{\gamma}}$. We find a considerably better sensitivity compared to the muon channel, with the $5\sigma$ discovery reach extending down to $BR_{\gamma\bar{\gamma}}\simeq 3.5\times 10^{-4}$ (\ie, roughly a factor 2 better than in the leptonic channel), and exclusion at 95\% CL for 
$BR_{\gamma\bar{\gamma}}\simeq 0.5\times 10^{-4}$ (\ie, about a factor 4 better than in the leptonic channel).

\begin{table}
\begin{center}
\begin{tabular}{|l||c|c|c|c|c|}
\hline
Process & Basic cuts &  $M_{jj}$ cut & $M_{\gamma\bar{\gamma}}$ cut & $M_{\rm miss}$ cut & $\slashed{E}$ cut \\
\hline
$jj  \gamma\bar{\gamma}$\;\; ($BR_{\gamma\bar{\gamma}} = 0.1\%$) & 804 & 669 & 154 & 110 & 72 \\
\hline
$jj\gamma$ & $3.39\times 10^{7}$ & $2.26\times 10^{7}$ & $1.47\times 10^{5}$ & $6.5\times 10^{4}$ & -- \\
\hline
 $jj\nu\bar{\nu}\gamma$ &  $3.9\times 10^{4}$  &  $3.1\times 10^{4}$   &  $5.9\times 10^{3}$   &  2.2   &  --  \\
\hline
\end{tabular}
\caption{Event yields after sequential cuts described in the text for $e^+e^-\!\!\rightarrow ZH\rightarrow q\bar q \gamma\bar{\gamma}$, and corresponding backgrounds, for an integrated luminosity of $10\ {\rm ab}^{-1}$, and  c.m. energy  $\sqrt{s} = 240\ {\rm GeV}$. The signal yield has been normalised assuming $BR_{\gamma\bar{\gamma}}=0.1\%$. Dashes stand for event yields less than 1.} 
\label{hadron channel cut flow}
\end{center}
\end{table}      

\begin{figure}
\begin{center}
\includegraphics[width = 0.7\textwidth]{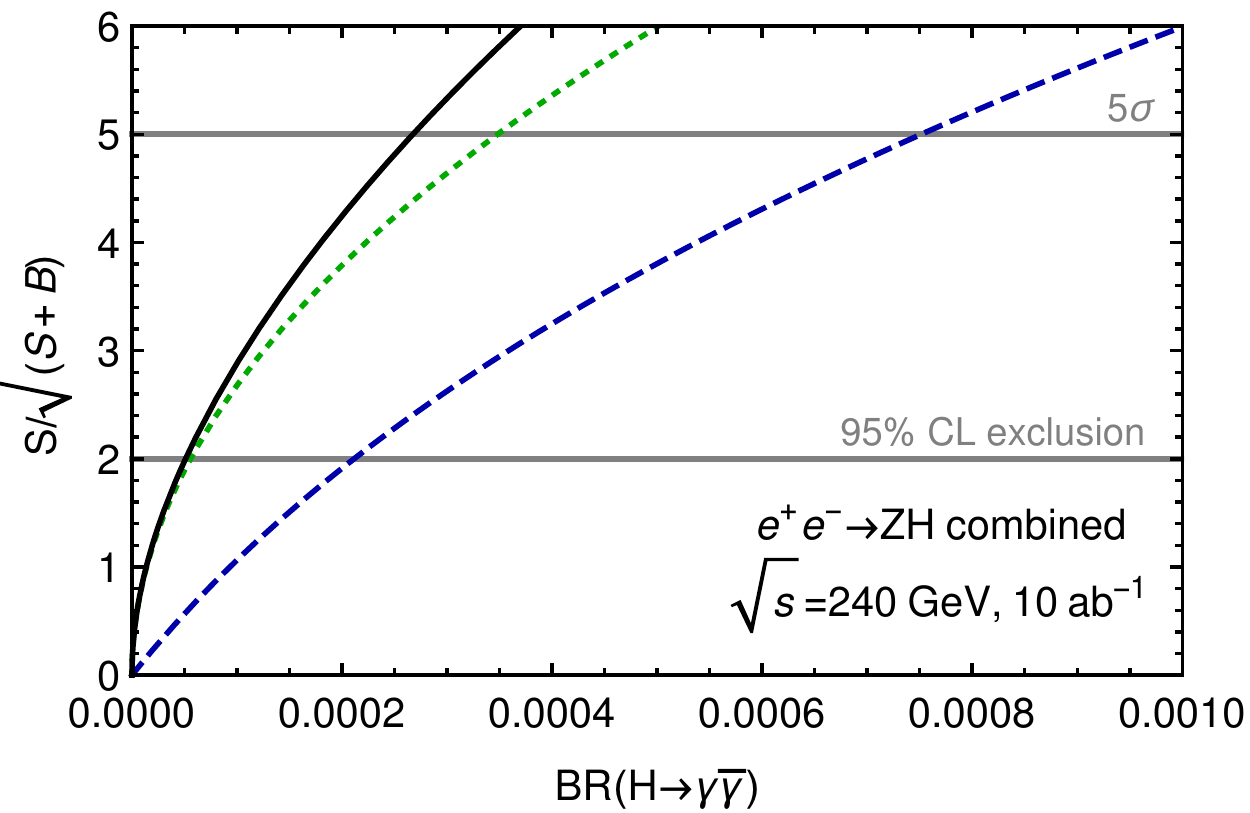}
\caption{Signal significance in the  $e^+e^-\!\!\rightarrow ZH\!\rightarrow \!q\bar q  \gamma\bar{\gamma}$ channel (green dotted line), $e^+e^-\!\!\rightarrow ZH\rightarrow \!\mu^+\mu^- \gamma\bar{\gamma}$ channel (blue dashed line) and in the combined search (black solid line) versus $BR_{\gamma\bar{\gamma}} $ for \mbox{10 ab$^{-1}$} at $\sqrt s=240$ GeV. The lower and upper horizontal lines pinpoint, respectively, the 95\% CL  exclusion bound, and the 5$\sigma$-significance discovery reach.
}
\label{combined significance}
\end{center}
\end{figure}

Finally, in Figure~\ref{combined significance}, we present the combined significance for the leptonic and hadronic searches. The combined $5\sigma$ sensitivity for discovery reaches  $BR_{\gamma\bar{\gamma}} \simeq 2.7\times 10^{-4}$, while the 95\%~CL exclusion reach is dominated by the hadronic channel sensitivity, and is again $BR_{\gamma\bar{\gamma}}\simeq 0.5\times 10^{-4}$.

\section{Conclusions}

A class of models potentially explaining the observed fermion mass hierarchy 
may naturally predict the decay of the Higgs boson into a photon and a dark photon
$\bar\gamma$ which is massless and undetectable by collider experiments. Thanks to the 
nondecoupling properties of the Higgs boson, the corresponding  branching ratio can be up to a few percent.

We have studied the potential of high-energy $e^+e^-$ facilities to either discover 
the $H\to \gamma \bar\gamma$ decay or
constrain its branching ratio.
In particular, we  have analysed the process $e^+e^-\rightarrow HZ$ followed by $H\rightarrow \gamma\bar{\gamma}$, considering both the leptonic channel where 
$Z\to \mu^+\mu^-$ and the hadronic channel where $Z\to q\bar q$, in $e^+e^-$ collisions  with integrated luminosity \mbox{10 ab$^{-1}$} at $\sqrt s\simeq 240$~GeV. In this setup,  the  production of about 2 million Higgs bosons is foreseen. 
We included initial-state radiation effects typical of a circular collider,
shower effects for the jet final states, and detector resolutions as presently foreseen for ILC detectors.

 We find that both the leptonic and hadronic $Z$ decay modes considerably contribute to the $e^+e^-\rightarrow ZH$ sensitivity, 
 with a quite higher potential for the hadronic mode.
 We have not analysed the $Z\to e^+e^-$ mode, which is expected to suffer from larger backgrounds and worse detector resolution with respect to $Z\to \mu^+\mu^-\!$.
 
 Discovery of the $H\to \gamma \bar\gamma$ decay with a $5\sigma$ sensitivity is reached 
 in $e^+e^-\rightarrow ZH$ for a branching ratio $BR_{\gamma\bar{\gamma}} \approx 2.7\times 10^{-4}$  by combining both muon and hadronic channels, while the corresponding 
 95\% CL exclusion  reach   is at  
\mbox{$BR_{\gamma\bar{\gamma}}\simeq 0.5\times 10^{-4}$}.
 
Note that this exclusion reach is more than two orders of magnitude better than the corresponding reach  of   the process \mbox{$e^+e^-\rightarrow H\bar{\gamma}$}  analyzed in~\cite{Biswas:2015sha}. 
On the other hand, the $e^+e^-\rightarrow ZH$ $5\sigma$ discovery reach is more than three times better than the LHC
reach with 300 fb$^{-1}\!$, and  comparable to the HL-LHC expected sensitivity, according to the preliminary analysis in~\cite{Biswas:2016jsh}.
Hence, the $e^+e^-\rightarrow ZH$ channel at FCC-ee/CEPC provides  a 
particularly sensitive probe to the Higgs branching ratio into a photon plus dark photon.

 We stress that this analysis  is model independent, and its results can be universally
applied to the search of any Higgs two-body decay into a photon plus an undetected
light particle, under the assumption of a SM $e^+e^-\rightarrow ZH$ cross section.
A modified Higgs production cross section can anyway be independently rescaled from our results.

Before concluding we note that the present analysis does not include  
{\it machine induced} backgrounds. In particular, 
beamstrahlung can considerably affect the impact of 
selection cuts in our signal-over-background optimisation strategy, by broadening the collision c.m. energy distribution.
On the other hand, beamstrahlung  is very much dependent on the actual accelerator technology, and 
 circular machines are much less affected by beamstrahlung with respect to linear colliders.  In fact, this potentially relevant effect 
can be accurately described  only after the basic machine parameters (and a particular scheme for beam bunches) will be set up (see for instance \cite{Voutsinas}). We anyhow think that the inclusion of such {\it machine induced} backgrounds is  beyond the scope of the present study.

\section*{Acknowledgements}
EG would like to thank the CERN Theoretical Physics Department for its kind hospitality during the 
preparation of this work. 
The work of MH has been supported by the Academy of Finland, grant 267842.

%

\end{document}